\documentclass[twocolumn]{aastex61}
\usepackage{amsmath}

\shorttitle{Time-variable star-planet interaction}
\shortauthors{Fischer \& Saur}

\begin{document}

\title{Time-variable electromagnetic star-planet interaction: \\
The TRAPPIST-1 system as an exemplary case}

\correspondingauthor{Christian Fischer}
\email{cfisch12@uni-koeln.de}

\author{Christian Fischer}
\affil{University of Cologne \\
Institute of Geophysics and Meteorology \\
Cologne, Germany}

\author{Joachim Saur}
\affil{University of Cologne \\
Institute of Geophysics and Meteorology \\
Cologne, Germany}



\begin{abstract}
Exoplanets sufficiently close to their host star can in principle couple electrodynamically to the star. This process is known as electrodynamic star-planet interaction (SPI). The expected emission associated with this coupling is however difficult to observe due to the bright intrinsic stellar emission. Identification of time-variability in the stellar lightcurve is one of the most promising approaches to identify SPI. In this work we therefore systematically investigate various mechanisms and their associated periods, which generate time-variability to aid the search for SPI. We find that the synodic and half the synodic rotation periods of the stars as measured in the rest frames of the orbiting exoplanets are basic periods occurring in SPI. We apply our findings to the example of TRAPPIST-1 with its seven close-in planets for which we investigate the possibility of SPI and the associated time-variabilities. We show that especially TRAPPIST-1b and c, are very likely subject to sub-Alfv\'{e}nic interaction, a necessary condition for SPI. Both planets are therefore expected to generate Alfv\'{e}n wings, which can couple to the star. The associated Poynting fluxes are on the order of $10^{11}$ to $10^{15}$ W and thus can hardly be the direct source of currently observable time-variability from TRAPPIST-1. However these Poynting fluxes might trigger flares on the star. We find correlations between the observed flares and the expected planetary induced signals, which could be due to SPI but our findings are not conclusive and warrant further observations and modelling.
\end{abstract}




\section{Introduction}\label{sec:intro} 
Electromagnetic star-planet interaction (SPI) generally describes the coupling between a planet and its host star. This coupling is in principle possible since planets are embedded in a magnetised plasma, i.e., a stellar wind, coming from the host star. In case the planet is sufficiently close to the star, the stellar wind conditions are such that the most important plasma wave, the Alfv\'{e}n wave, can travel upstream and thus establish an electromagnetic coupling between the planet and the star. This condition is met if the Alfv\'{e}n wave velocity is larger than the stellar wind velocity, which is referred to as sub-Alfv\'{e}nic. The resultant structure of the Alfv\'{e}n wave is called Alfv\'{e}n wing. It can carry large energy fluxes and is well known from, e.g, Jupiter and its moons. The Jupiter-Io interaction has been studied for decades, e.g., with the unipolar inductor model by \citet{GolreichLyndenBell69} and the Alfv\'{e}n wing models by \citet{Neubauer80} and \citet{Goertz80}.

The main concept of the Alfv\'{e}n wing model is that the planet slows down the plasma around it, e.g., through mechanical interaction by collisions with particles of the neutral atmosphere \citep{Neubauer80}. The magnetic field is partly frozen into the plasma and builds up magnetic tensions, caused by the plasma deceleration. These perturbations propagate along the field lines as Alfv\'{e}n waves. The models by \citet{Lanza09,Lanza12,Lanza13} are alternative approaches to SPI. The models assume magnetohydrostatic force-free magnetic fields coupling the planet and the star. In \citet{Lanza09} the interaction drives an electric current system that closes through reconnection near the planetary magnetopause. \citet{Lanza13} propose another model based on the the accumulation and release of stresses in magnetic flux tubes driven by the motion of the planet through the stellar magnetic field. For a review of the different models we refer the reader to \citet{Saur17}. \citet{Zarka07} derived scaling laws for radio emissions emanating from SPI. The author differentiates between the interaction of magnetised and non-magnetised planets with the surrounding stellar winds based on the moon-planet interactions at Jupiter. In this study we apply the Alfv\'{e}n wing model because it is a well established concept that describes the moon-planet interaction in our solar system and has been applied in the exoplanetary context due to expected physial similarities \citep{Saurea13,Strugarek16}.

SPI has also been investigated in a variety of numerical studies. \citet{Ipea04} investigated the local interaction of exoplanetary magnetospheres under sub-Alfv\'{e}nic conditions. \citet{Preusseea05,Preusseea06,Preusseea07} investigated Alfv\'{e}n characteristics and the local sub-Alfv\'{e}nic interaction of Hot Jupiter magnetospheres. \citet{Koppea11} simulated the systems HD 179949 and $\upsilon$ And and confirmed the theoretical explanation for the phase shift between planet and observed emission. \citet{Strugarekea12,Strugarekea14,Strugarekea16} developed a numerical model to quantify the angular momentum transfer between star and planet and described the impact of electromagnetic SPI on planet migration. \citet{Cohenea09,Cohenea11,Cohenea14,Cohenea15} investigated the interaction of planetary magnetospheres with the host star and the impact of stellar winds on planets.

Observational evidence for SPI exists for a handful of systems, notably HD 179949 \citep{Shkolnikea03,Shkolnikea05,Shkolnikea08} and $\mathrm{\upsilon}$ And \citep{Shkolnikea05, Shkolnikea08}. The authors observed enhanced \ion{Ca}{2} H \& K emission in synchronisation with the orbital period of the close-in exoplanet around HD 179949 in four of six observational epochs. They also saw similar effects at $\upsilon$ And. \citet{Shkolnikea08} account for this on-off nature of the interaction by invoking a changing stellar magnetic field structure. \citet{Staabea17} observed the system WASP-43 and likewise saw enhanced \ion{Ca}{2} H \& K emissions, which they attributed to the existence of SPI. \citet{Walkerea08} observed the star $\tau$ Boo A photometrically and in the chromospheric \ion{Ca}{2} K line. The star rotates tidally locked to its planet and features a variable region in advance of the planet that is attributed to magnetic SPI \citep{Walkerea08}. \citet{Millerea12} observed the system WASP-18 in optical and X-ray wavelength ranges and see no evidence for SPI. The authors claimed WASP-18 b to potentially generate the strongest interaction among the known planets at that time, based on planetary mass and semimajor axis. X-ray observations show that WASP-18 appears to be less active than expected from its estimated age \citep{Millerea12}. According to \citet{Millerea12} the lack of activity indicates a much weaker stellar magnetic field than in HD 179949, which directly affects the strength of potential SPI. \citet{Millerea12} also show by the examples of HD 179949 and $\mathrm{\upsilon}$ And that a chromospheric hot spot might mimic the existence of SPI. \citet{Saarea08} observed a possible link between stellar X-ray emission and a close-in planet at HD 179949. X-ray observations of $\upsilon$ And conducted by \citet{Poppenhaegerea11} provided evidence of activity linked to stellar rotation but no connection to the planet. \citet{Scharf10} saw correlations between stellar X-ray luminosities and orbiting planets. Another statistical study by \citet{Poppenhaegerea10} however did not find any correlation between X-ray and the existence of planets. \citet{PoppenhaegerWolk14} found evidence for enhanced activity of a Hot-Jupiter hosting binary compared to binaries without Hot Jupiters. Another line of searches for SPI comes from H-$\alpha$ and radio observations of the ultracool dwarf LSR J1835+3259  by \citet{Hallinanea15}. The authors suggest that the emission contains auroral signatures similar to those on Jupiter possibly caused by a planetary companion. However, observations with the Hubble Space Telescope reported in \citet{Saurea18} show that the UV spectrum of LSR J1835+3259 rather resembles that of late M-dwarf stars instead of auroral emissions similar to those from Jupiter. However, although there are quite intriguing observations for certain systems, many stellar systems that are expected to host electromagnetic SPI lack convincing evidence (see review by \citet{Lanza15}). Independent approaches based on different observations are therefore warranted to better investigate and possibly identify SPI in these systems.

TRAPPIST-1 and its planets are an intriguing system for our studies. TRAPPIST-1 is a star of spectral class M8, lies at a distance of 39 Ly, and has a mass of 84 Jupiter masses, but is only approximately as large as Jupiter \citep{Gillonea16}. TRAPPIST-1's very low mass implies that it is close to the border of the Brown Dwarf regime. \citet{Gillonea16} carried out optical spectroscopy and did not detect the 670.8 nm lithium line. This implies that TRAPPIST-1 belongs to the very low mass stars, instead of the Brown Dwarfs. This classification is supported by measurements of TRAPPIST-1's X-ray and UV-emissions \citep{Bourrierea17,Wheatleyea17}, which indicate the existence of a hot stellar corona and a moderately active Chromosphere. \citet{Gillonea16, Gillonea17} discovered seven close-in, terrestrial planets in the TRAPPIST-1 system. The planets orbit TRAPPIST-1 at distances between $20$ and $109$ stellar radii $R_\mathrm{*}$ where three to four of them are believed to lie in the habitable zone \citep{Gillonea16,OMalleyJamesea17}. Due to this large number of potentially habitable planets several telescopes carried out further observations of the system, including K2 \citep{Vidaea17,Lugerea17}, Spitzer and the TRAPPIST-telescope \citep{Gillonea17}. While the orbital periods of the planets are well known, there is some controversy about the stellar rotation period $T_*$. \citet{Lugerea17} and \citet{Vidaea17} proposed $T_*$ to be 3.3 d based on periodicities observed by K2. This is in contrast to earlier measurements by \citet{ReinersBasri10} and \citet{Gillonea16} who reported a rotation period of around $1$ d. \citet{Reinersea18} conducted more precise radial velocity measurements and concluded a period of $T_* \, \sin \ i > 2.7$ d, whereas the inclination $i$ is unknown. These new measurements are consistent with the rotation period obtained by \citet{Lugerea17} and \citet{Vidaea17}.

Observation of temporal variability associated with SPI might be one of the most promising ways to detect SPI. In this paper we therefore investigate how SPI can be identified based on its temporal variabilty. We start this work with basic theoretical considerations of SPI, from which we will derive different mechanisms of temporal variability and their effects on stellar lightcurves. In the second part of this paper we choose the example of TRAPPIST-1 to investigate the SPI of its planets and their potential observability due to time-variability. Therefore we carry out theoretical studies on the existence and possible properties of SPI generated by the seven planets in the TRAPPIST-1 system. Then we compare our results with the recent K2 observations to see if there are hints of temporal variability induced by SPI.

\section{Identification of time-variable SPI in stellar lightcurves}\label{sec:methodsection} 
In this section we will investigate mechanisms that cause temporal variability in SPI. The identification of signals that appear with periods connected to these mechanisms could help to detect SPI in photometric lightcurves.

As described in section \ref{sec:intro} the planets have to experience sub-Alfv\'{e}nic conditions to generate Alfv\'{e}n wings. Therefore the local Alfv\'{e}n Mach number
\begin{equation}
    M_\mathrm{A}=\frac{v_\mathrm{rel}}{v_\mathrm{A}}\label{eq:Machnumber}
\end{equation}
has to be smaller than one. Here $\textit{\textbf{v}}_\mathrm{rel}=v_\mathrm{sw}\,\textit{\textbf{e}}_r-v_\mathrm{orb}\,\textit{\textbf{e}}_\varphi$ is the relative velocity between the stellar wind and the planet that moves with Keplerian velocity $v_\mathrm{orb}$, and $v_\mathrm{A}=B\,\sqrt{\mu_0\,\rho}\,^{-1}$ is the Alfv\'{e}n velocity, with the magnetic field $B$ and the mass density $\rho$. The vector $\textit{\textbf{e}}_r$ is the unit vector in radial direction and $\textit{\textbf{e}}_\varphi$ the unit vector in orbital direction of the planet. The Alfv\'{e}n wing carries electromagnetic energy towards the star. The energy flux is given by the Poynting flux and was derived for small $M_\mathrm{A}$ by \citet{Saurea13} as
\begin{equation}
   S=2\pi R_\mathrm{eff}^2\,\frac{(\bar{\alpha}\,M_\mathrm{A}\,B_\mathrm{loc}\,\cos \theta)^2}{\mu_0}\,v_\mathrm{A}.\label{eq:poyntingformula}
\end{equation}
$R_\mathrm{eff}$ is the effective size of the obstacle, $B_\mathrm{loc}$ the local magnetic field strength and $\theta$ the angle between magnetic field normal and the relative velocity $\textbf{v}_\mathrm{rel}$. The interaction strength $\bar{\alpha}$ takes values between $0$ and $1$. Sophisticated MHD-simulations of star-planet interaction by \citet{Strugarekea15} demonstrate that the analytic expression of equation \ref{eq:poyntingformula} reproduces the numerically modelled values of the Poynting fluxes remarkably well.

\subsection{Mechanisms for temporal variability in SPI}\label{sec:tempvarmech}
Star-planet interaction is hard to observe because the generated power is expected to be much weaker than the bolometric stellar emission \citep{Saurea13}. The equivalent to SPI from our solar system, the interaction between the giant planets and their moons, is comparably simple to observe. Emissions of the planetary aurora including the moons' auroral footprints can be spatially resolved in observations \citep{Clarkeea02,Pryorea11}. In contrast the emissions generated by SPI might be hard to distinguish from the stellar background but might be visible in certain parts of the stellar spectrum, i.e., typical coronal or chromospheric emissions like UV, X-ray or H$\alpha$ and \ion{Ca}{2} H \& K. Knowledge about the expected periodicities of time-variable SPI helps to identify planet-related emissions in observations. Stars usually have an intrinsic variability that is often connected to corotating features in their atmospheres and therefore allow to estimate the stellar rotation period. In addition to that, there is a certain randomness in the occurrence of flares, prominences and coronal mass ejections on stars. \citet{Lanza18} investigated how SPI can trigger flares. Therefore the periodic occurrences of flares might indicate the existence of electromagnetic star-planet interaction in a stellar system.

The most basic periods that occur in stellar systems are the rotation period of the star $T_*$ and the orbital period of its planet $T_P$. Both are the respective sideric periods relative to a fixed star. In multi-planet systems of course there are also multiple orbital periods.
Temporal variability of the SPI can come from both of these basic periods in a system. The reason is that the Alfv\'{e}n wing's footpoint moves with the planet's orbital period across the stellar surface. 
The strength of the SPI is controlled by properties of the rotating star and properties of the orbiting planet. This combination results in a periodic time-variability with the synodic rotation period of the star as seen from the planet. The synodic periods generally are given by 
\begin{equation}
   T_\mathrm{syn}=\left|\frac{2\pi}{\Omega_\mathrm{inner}-\Omega_\mathrm{outer}}\right|\label{eq:Tsyn}
\end{equation}
where $\Omega$ and  $T=2\,\pi/\Omega$ are the angular velocities and the sideric rotation periods, respectively. The subscript $inner$ stands either for the stellar rotation or an inner planet in case of an interaction between two Alfv\'{e}n wings. The subscript $outer$ stands for the corresponding outer planet.

In our considerations about temporal variability we make certain assumptions that allow an analytic treatment of these processes. We assume that the planet moves on a circular orbit, which likely holds well for close-in planets due to the tidal interaction with the star, just as it is the case in the TRAPPIST-1 system \citep{Gillonea17}. Further we assume a dominating dipolar magnetic field and a steady state stellar wind.

\begin{figure*}
\plotone{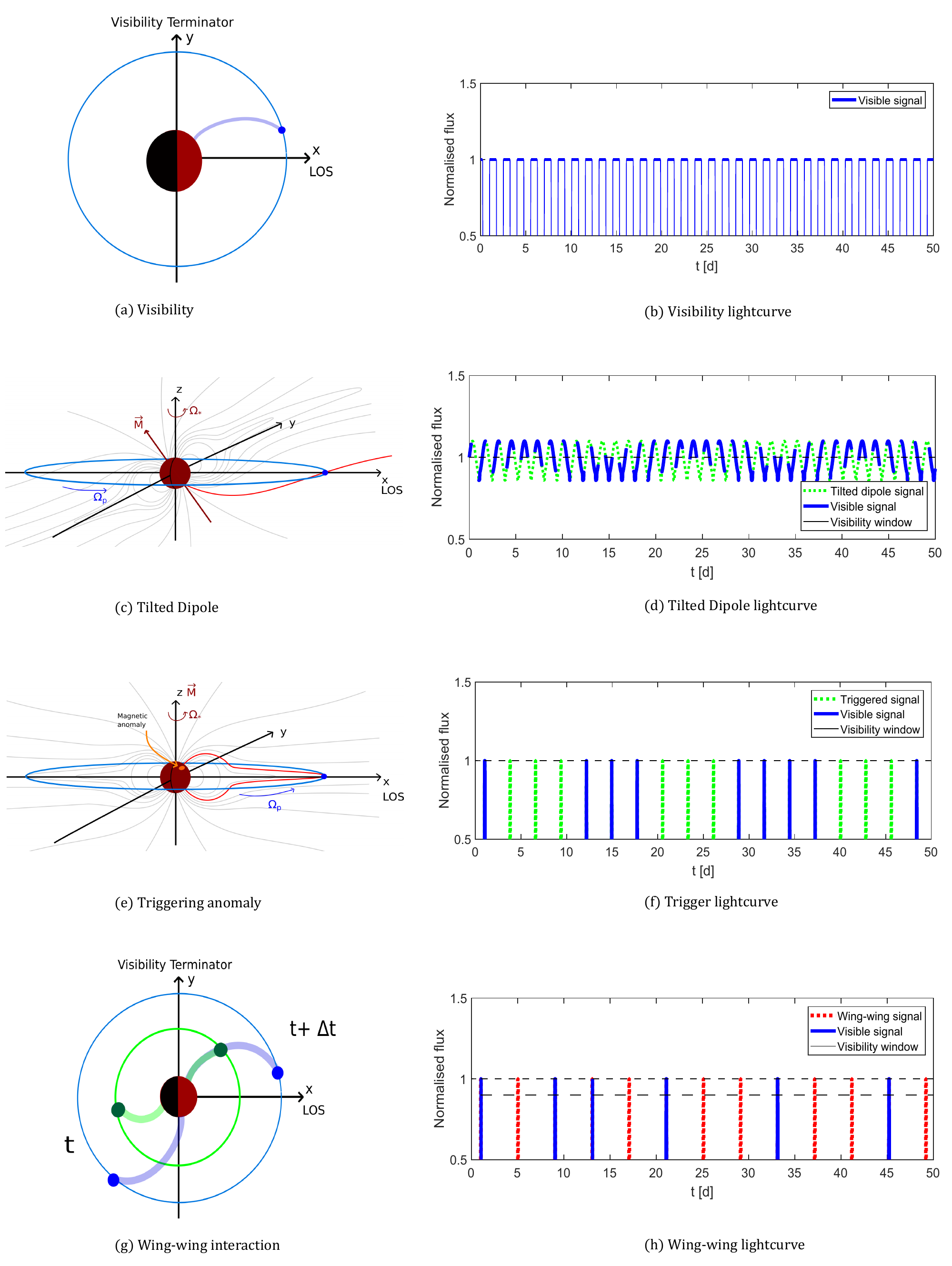}
\caption{The figures show sketches and expected lightcurves for four different mechanisms, which can cause temporal variability in SPI. (a) and (b) show the visibility of the signal, (c) and (d) the effect of a tilted dipole, (e) and (f) depict the effect of a magnetic anomaly region on the star and (g) and (h) show how wing-wing interaction, i.e. planet-planet interaction might look like.\label{fig:MechanismsandLightcurves}}
\end{figure*}

Under these conditions we identified four mechanisms that could cause a temporal variability in SPI. These mechanisms are the visibility of the Alfv\'{e}n wing, effects related to a tilted dipole magnetic field, a magnetic anomaly on the star or the interaction of multiple Alfv\'{e}n wings with each other. All of these causes for time-variability are presented in figure \ref{fig:MechanismsandLightcurves}. The corresponding panels show a sketch of the physical situation on the left and a corresponding lightcurve of the generated signal on the right. All related periods here are shown for the special case of TRAPPIST-1 and its planets T1b and T1c. These processes introduce three qualitatively different periods for the SPI: The orbital period of the planet $T_\mathrm{p}$, the synodic period between stellar rotation and planetary orbit $T_\mathrm{syn}$ and the synodic period between two planets $T_\mathrm{syn}^\mathrm{planet}$. In the following sections we will discuss these different mechanisms, which produce temporal variability with their respective properties. Also we will outline in what way the star could respond to an interaction with Alfv\'{e}n wings and how this in turn affects the observability of the SPI.

\subsubsection{Visibility}\label{sec:visibility}
The visibility of the Alfv\'{e}n wing is the most basic process that causes a temporal variability. Due to the motion of the planet the Alfv\'{e}n wing moves across the star and is either on the visible hemisphere or on the other side. The purest realisation of this process is if there are no latitudinal asymmetries altering the signal of the Alfv\'{e}n wing.

Figure \ref{fig:MechanismsandLightcurves} (a) shows a top view of the physical situation with the x-axis pointing into the line of sight (LOS) towards the observer and the z-axis of the coordinate system being aligned with the stellar spin axis. The shaded region on the star represents the non-visible hemisphere where the y-z-plane determines the visibility terminator. The planet orbits the star and creates the signal in figure \ref{fig:MechanismsandLightcurves} (b). This shows the hypothetical normalised lightcurve over a time of 50 d and without any intrinsic luminosity coming from the star. We applied the periods of TRAPPIST-1 and its innermost planet as an example to demonstrate this process. The period of the signal is accordingly $T_\mathrm{SPI} = T_\mathrm{p}$ and has, in the simplest case of constant emission, the shape of a boxcar function. This boxcar function needs to be modified as a function of time depending on the optical depth of the emitted radiation through the stellar atmosphere

\subsubsection{Tilted dipole magnetic field}
The energy flux generated at the planet depends on the magnetic field, the plasma density and the velocity of the stellar wind. In a tilted dipole magnetic field all three parameters vary as a function of magnetic latitude. Thus the Poynting flux varies as a function of time. 

Assuming that stellar winds form similar structures as the solar wind there will be a fast wind originating from higher latitudes and a slow wind from lower latitudes. The slow wind zone also includes the stellar current sheet formed by a thin region of oppositely directed field lines. In the current sheet the magnetic field lines are closed and thus the field strength decreases stronger with radial distance than in the ambient stellar wind. In order to maintain pressure balance the plasma density in the current sheet is higher than in the surrounding stellar wind \citep{Smith01}.

Figure \ref{fig:MechanismsandLightcurves} (c) shows such a situation. The coordinate system is the same as in figure \ref{fig:MechanismsandLightcurves} (a). The sketch shows a tilted stellar dipole field, represented by the magnetic dipole moment $\textit{\textbf{M}}$. We can see the region close to the star, where the lower-latitude field lines are closed dipole field lines. In polar regions the field lines are quasi-open because the stellar wind stretches them much beyond the orbit of the planets. Current sheet field lines are stretched in the outward/inward direction and originate on latitudes around the magnetic equator.

For the planet there are two hypothetical situations. The one presented in figure \ref{fig:MechanismsandLightcurves} (c) shows the planet located on quasi-open field lines. In the second situation the planet is constantly located within closed field lines, which only occurs for close-in planets.

The expected period of the occurring signal is $T_\mathrm{SPI} = T_\mathrm{syn}/2$ for both situations, i.e., half the synodic rotation period of the star in the frame of the planet. The planet crosses the slow wind zone twice during each rotation of the star, which creates the quasi-periodic SPI signal with half the synodic period. In the slow wind zone and the current sheet the stellar wind properties (such as local density and magnetic field strength) change and affect the amplitude of the Poynting flux generated by the planets (see expression \ref{eq:poyntingformula}), which is the physical reason for temporal variability in SPI. Especially if the planet resides on quasi-open field lines the differences to the current sheet might be strong. The resulting qualitative signal could have a shape like the green curve in figure \ref{fig:MechanismsandLightcurves} (d). The figure shows the relative incoming energy flux of the star-planet interaction. If we further take visibility into account we obtain the blue curve. We indicated the visibility windows with black bars, appearing with a period of $T_\mathrm{p}$ (in this case 1.51 d) and a duration of $T_\mathrm{p}/2$. Just as in section \ref{sec:visibility}, these windows indicate the times when the Alfv\'{e}n wing's footpoint resides on the visible hemisphere of the star. This effect breaks the periodicity of $T_\mathrm{syn}/2$ to some extent. Instead we see a beat-like interference pattern with varying powers and a much larger period.

This mechanism is well known from the Jupiter system. Due to the tilt in Jupiter's magnetic field, the jovian moons experience time-variable magnetic field strengths and plasma densities, since the bulk of the plasma is concentrated in the centrifugal equator \citep{HillMichel76,Bagenalea80}. The Io footprint undergoes continuous brightness variations \citep{Wannawichianea10,Wannawichianea13} caused by variable Poynting fluxes with a period of $T_\mathrm{syn}/2$ \citep{Saurea13}.

\subsubsection{Triggering Anomaly}\label{sec:TriggerAnomaly}
The idea behind the trigger mechanism is that the planet's Alfv\'{e}n wing releases energy, for example stored in form of magnetic loops, on the star. It has been proposed by \citet{Lanza09,Lanza12,Lanza13} as a change in magnetic helicity, which releases free magnetic energy. This energy release causes a flare event with a significantly larger energy output compared to energy deposition of the Alfv\'{e}n wing energy flux during a comparable time. \citet{Lanza18} describes three different mechanisms that result in the release of energy stored in magnetic loops and provides relations to estimate upper limits of the flare energies. For two of these mechanisms the planet may act as a trigger for the eruption of a flare. However flares can also erupt without the influence of a planet. According to \citet{Lanza18} both processes create flare energies of about $10^{26}$ J up to $10^{31}$ J, depending on the star. The third described mechanism only operates in the presence of a planet. The planet's interaction with an open coronal magnetic field generates the flare. The created energies are typically one to three orders of magnitude smaller than for the other two mechanisms.

Figure \ref{fig:MechanismsandLightcurves} (e) depicts the simplest case of a large scale dipole field with an anomaly indicated by the orange dot on the star. The Alfv\'{e}n wing, indicated in red, intersects this anomaly and triggers the release of energy. This situation causes a periodicity of $T_\mathrm{SPI} = T_\mathrm{syn}$ if the triggered signal is much stronger than the power of the plain Alfv\'{e}n wing.
The green curve in figure \ref{fig:MechanismsandLightcurves} (f) shows the expected signal of this mechanism.  If we take visibility into account we receive the signals shown in blue. As for the tilted dipole the inclusion of visibility breaks the overall periodicity. We only see piecewise periodic signal peaks whereas there are large intervals with no signal.

\begin{figure*}
\gridline{\fig{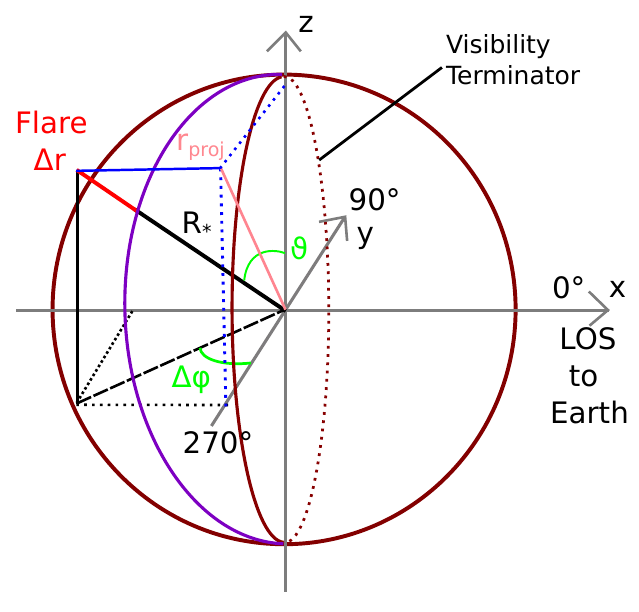}{0.4\textwidth}{}}
\caption{This figure shows the geometry of the extended visibility when a flare with height $\Delta r$ erupts behind the visibility terminator. The $\varphi=0^\circ$ longitude points into the LOS and $\vartheta$ is the colatitude.} \label{fig:ExtendedVisibility}
\end{figure*}

However the spatial extent of a flare will extend the visibility range to more than the visible half-sphere. Figure \ref{fig:ExtendedVisibility} (a) shows a simplified geometry that allows the estimation of an extended visibility surface due to the elevation of flares. The coordinate system is the same as in figure \ref{fig:MechanismsandLightcurves}, with the x-axis pointing towards the Earth and the z-axis being aligned with the spin axis of the star. The great circle in the y-z-plane is therefore the visibility terminator.

We assume that a flare ejects radially outward with a height $\Delta r$ (figure \ref{fig:ExtendedVisibility} (a) bright red line) and is located on a colatitude $\vartheta$ and at an azimuthal position behind the visibility terminator $\Delta \varphi$ (both indicated in green). If such a flare is high enough that its top is visible above the visible disk, the flare's footpoint belongs to the range of extended visibility.
We aim to find the longitudinal extent $\Delta \varphi$ into the non-visible side such that the top of the flare is still visible for a given flare height and flare latitude. In observations we see the top of the flare projected into the $y$-$z$-plane, which is given by $r^2_\mathrm{proj}=y^2+z^2$ (pink) with $y=(R_*+\Delta r)\,\sin\vartheta\,\cos\Delta\varphi$ and $z = (R_*+\Delta r)\,\cos\vartheta$. We use $r^2_\mathrm{proj}>R_*^2$ and receive 
\begin{equation}
\sin \Delta \varphi < \frac{\sqrt{1-\left(\frac{R_*}{R_*+\Delta r}\right)^2}}{\sin \vartheta} \label{eq:VisExtensionFormula}
\end{equation}
which describes the maximum extension of the flare's visibility. The visible hemisphere covers 50\% of the stellar surface and the extended visibility reaches up to
\begin{equation}
\mathrm{Vis} = \frac{180^\circ + 2\Delta \varphi}{360^\circ}\cdot100\%.
\end{equation}
We note that this effect is more relevant for larger ratios of flare height to stellar radius. It is apparent from equation \ref{eq:VisExtensionFormula} that the visibility will be 100\% for flares appearing at the poles, whereas the minimum of the visibility extension will be at the stellar equator.

\citet{Parker88} proposed the existence of nanoflares with an energy $<10^{20}$ J on the sun. The main idea of this work was that the corona is heated by nanoflares that appear on large scales. \citet{Aschwandenea2000} proposed a classification that divides solar flare events by their total energy with microflares being below $10^{23}$ J and nanoflares below $10^{20}$ J. The same study investigated the occurrence and typical emissions of different types of flares. High energy flares that emit hard X-rays are accordingly rare events. The flare frequency increases with decreasing flare energy. Nanoflares are about $10^{5}$ to $10^{10}$ times more frequent than high energy flares. Microflares tend to emit soft X-rays while nanoflares emissions lie in the EUV range. An Alfv\'{e}n wing might trigger a large number of these smaller events along its path across the star. Assuming quasi-constant flare triggering we expect those emissions to be pulsed with the period of the planet's orbit similar to the discussed effects in section \ref{sec:visibility}. Also these events might account for the enhanced X-ray activity in some planet hosting stars. However strong flares in contrast might occur when regions with active magnetic loops are triggered by an Alfv\'{e}n wing.

\citet{Lanza09,Lanza12,Lanza13} and \citet{Saurea13} discussed the relevance of a trigger mechanism in the case of HD 179949 \citep{Shkolnikea03,Shkolnikea05}. Both SPI models predict powers of about $10^{17}$ W that are generated by the planet, while the observations from \citet{Shkolnikea05} indicate excess powers of $10^{20}$ W, which lead to the introduction of the trigger mechanism. Another aspect is that \citet{Shkolnikea03,Shkolnikea05} observed emissions that are pulsed with the planetary orbital period. As discussed in this section we would expect the signal to appear with the respective synodic rotation period of the star as seen from the planet if flares are triggered at a fixed longitude on the star. In the case of HD 179949 the synodic period is approximately $4.5$ d, considering the planetary orbital period of $3$ d \citep{Shkolnikea03} and the stellar rotation period of $7$ to $9$ d \citep{Shkolnikea08}. In case of triggering of flares across the surface of the star on time scales much smaller than the rotation period \citep{Lanza09}, the resultant lightcurve will be dominated by the sideric rotation period of the planet.

\subsubsection{Wing-wing interaction}
Alfv\'{e}n wings exist in Jupiter's magnetosphere, where the magnetic field can approximately be described as a dipole field. Therefore the Alfv\'{e}n wings generated by each moon follow the field lines that are connected to the moon into the polar jovian ionosphere. This implies that Alfv\'{e}n wings of different moons never interact with each other. 

The situation could be different in stellar environments with SPI. Sufficiently far from the star, the magnetic field geometry differs from the dipole field lines around planets because the stellar magnetic field lines are carried out with the stellar wind. This effect leads to extremely stretched current sheet magnetic field lines, which are very close to the spin plane of the star. If the planets reside in the current sheet and two planets coincidently share the same quasi-open field line, the Alfv\'{e}n wings of those planets might intersect and interact with each other. This interaction might mutually affect the two planets that are involved, e.g., by enhanced auroral activity. Therefore this interaction could alternatively be named planet-planet interaction. The idea is sketched in figure \ref{fig:MechanismsandLightcurves} (g). The planets would have to lie within the current sheet while the field lines have to be sufficiently parallel to each other, in order to allow the Alfv\'{e}n wings to merge. On the left there is a hypothetical situation at time $t$ where both Alfv\'{e}n wings are separated from each other and later at time $t+\Delta t$ both Alfv\'{e}n wings merge into each other.

Wing-wing interaction has not received attention in the literature to the authors' knowledge. The interaction of Alfv\'{e}n wings generates non-linear effects, which might result in an intensification of a possible joint wing. The wing-wing interaction occurs at the difference of the orbital angular velocities of both planets, i.e., the synodic rotation period of one planet as seen in the rest frame of the other planet $T_\mathrm{SPI} = T_\mathrm{syn}^\mathrm{planets}$. The signal is indicated in red in figure \ref{fig:MechanismsandLightcurves} (h) on the example of T1b and T1c. To be visible the footprint of the wing-wing interaction has to lie on the visible hemisphere. As one can see in the blue curve this clearly breaks any periodicity of the signal.

\section{SPI in the TRAPPIST-1 system}
For our study of SPI and its associated time-variabilites we chose the example of TRAPPIST-1 because of its large number of close-in planets, which give a larger number of possible periods to observe. We apply the semi-analytical model of TRAPPIST-1's stellar wind from appendix \ref{sec:AppStellarWindModel} to investigate with an appropriate parameter space if SPI is possible in this system and to identify the most likely planets that cause SPI. Therefore we calculate radial profiles of the Alfv\'{e}n Mach number via equation \ref{eq:Machnumber} and the Poynting flux with equation \ref{eq:poyntingformula}. Based on those results we will try to identify time-variable signals of SPI in the K2 lightcurve of TRAPPIST-1.

\begin{deluxetable*}{c|cc|c|CC}[t]
\tablenum{1}
\tablecaption{Results and parameters of TRAPPIST-1 and its seven planets: Orbital distance $a$, period $T$ and the estimated synodic periods of the interaction. The remaining columns show the best-guess results for the Mach number $M_\mathrm{A}$ and the Poynting flux $S$\label{tab:SPIresults}}
\tablewidth{0pt}
\decimals
\tablehead{
\colhead{Object} & \colhead{$a$ [$R_\mathrm{*}$]\tablenotemark{a}} & \colhead{$T$ [d]\tablenotemark{a}} & \colhead{$T_\mathrm{syn}$ [d]}& \colhead{$M_\mathrm{A}$\tablenotemark{b}} & \colhead{$S$ [$10^{13}$ W]\tablenotemark{b}}
}
\startdata
T1 & 1 & 3.3 &  & & \\
T1b & 20.50 & 1.51087  & 2.79  & 0.38 \ _{0.02}^{20} & 14.2 \ ^{233}\\
T1c & 28.08 & 2.42182  & 9.10  & 0.55 \ _{0.03}^{28} & 2.5 \ ^{4.3}\\
T1d & 39.55 & 4.0496   & 17.82 & 0.80 \ _{0.05}^{40} & 0.26 \ ^{0.43}\\
T1e & 51.97 & 6.0996   & 7.19  & 1.08 \ _{0.06}^{54} & 0 \ ^{2.5}\\
T1f & 68.40 & 9.2067   & 5.14  & 1.45 \ _{0.09}^{72} & 0 \ ^{4.3}\\
T1g & 83.20 & 12.3529  & 4.50  & 1.79 \ _{0.11}^{89} & 0 \ ^{1.6}\\
T1h & 109   & 18.767   & 4.00  & 2.58 \ _{0.15}^{128} & 0 \ ^{0.49}\\
\enddata
\tablenotetext{a}{\citet{Gillonea16,Gillonea17,Lugerea17}}
\tablenotetext{b}{See ranges in figures \ref{fig:Machnumbers} and \ref{fig:Poyntingfluxes} for upper and lower boundaries. The lower boundaries for S are zero.}
\end{deluxetable*}

\subsection{Parameter space}\label{sec:parameterspace} 
To characterise possible SPI we need to know the stellar wind and the magnetic field of TRAPPIST-1. Since very little is known about the winds of late M-dwarfs such as TRAPPIST-1, it is commonly assumed in the literature that these winds are similar to the solar wind and therefore thermally driven \citep{Cohenea14,Vidottoea11,Garraffoea16,Garraffoea17}. However only a few parameters which constrain its stellar wind have been observed for TRAPPIST-1, otherwise we take typical values that are known for other M-dwarfs. TRAPPIST-1 with its spectral class M8 belongs to the late-type M-dwarfs whereas most parameters are measured for early or mid M-dwarfs.

We chose the stellar wind model presented in appendix \ref{sec:AppStellarWindModel} in such a way that there are only three independent parameters: the coronal temperature $T_c$, the total mass flux $\dot{M}$ and the equatorial magnetic field $B_0$.
\citet{ReinersBasri10} estimated the mean magnetic field strength as $0.06$ T, with an uncertainty range from $0.02$ T to $0.08$ T. The geometry of the field is unknown. \citet{Garraffoea17} applied a magnetic field map with a strongly tilted dipole based on observations of the mid M-dwarf GJ 3622. Field geometries of late M-dwarfs are generally unknown because of the stars' faintness. \citet{Morin10} observed the M8 star VB10 and only detected a weak signal which allowed the conclusion that the field has an axisymmetric toroidal component. In general, the magnetic fields of late M-dwarfs are believed to become more axisymmetric and dipolar \citep{Morin10,ReinersBasri09}, but probably also display asymmetric higher-order magnetic multipole components \citep{Morin10}. \citet{Berdyuginaea17} measured the surface magnetic field strength of the ultracool dwarf LSR J1835+3259. In their measurements they identified a spot with strongly increased field strength. The authors were not able to determine the field geometry but the existence of this spot indicates multipole field components. In the absence  of further information for TRAPPIST-1, we simply use a pure dipole field as the source field for our stellar wind model.

TRAPPIST-1's coronal temperature $T_c$ has not been measured yet. For M-dwarfs, temperatures of $2$ to $3\cdot10^6$ K, similar to the solar coronal temperature, are applied in the literature \citep{Vidottoea14,Garraffoea16}. For TRAPPIST-1, \citet{Garraffoea17} chose a hot sun-like corona. \citet{Wheatleyea17} report an X-ray luminosity of TRAPPIST-1 that is similar to the quiet sun and indicates the existence of a hot stellar corona. Therefore we can expect temperatures of about $10^6$ K as in the solar corona. M-dwarfs generally also show evidence for very high coronal temperatures up to $10^7$ K \citep{Schmittea90,Giampapaea96}. Hence we choose our parameter space accordingly with $T_c$ in the range $10^6$-$10^7$ K.

The parameter $\dot{M}$ has been derived from observations for different M-dwarfs but not for TRAPPIST-1. \citet{Vidottoea14} infer a range between approximately $10^8\,\mathrm{kg}\,\mathrm{s}^{-1}$ and $10^{12}\,\mathrm{kg}\,\mathrm{s}^{-1}$ based on observations from \citet{Woodea01} and \citet{Mullanea92}. For Proxima Centauri the maximum $\dot{M}$ is estimated to be approximately $10^{10}\,\mathrm{kg}\,\mathrm{s}^{-1}$ \citep{WargelinDrake02}. \citet{Turnpenneyea18} and \citet{Garraffoea17} assume $2\cdot10^9\,\mathrm{kg}\,\mathrm{s}^{-1}$ for TRAPPIST-1, which corresponds to a sun-like mass outflow. The latter authors already assumed a similar value for their simulations of Proxima Centauri \citep{Garraffoea16}. On this basis we chose a parameter range of $10^8-10^{12}\,\mathrm{kg}\,\mathrm{s}^{-1}$. 

Additionally to the parameter space study we apply a 'best guess' with corresponding values of $\dot{M}=10^{10}\,\mathrm{kg}\,\mathrm{s}^{-1}$, $T_c = 2\cdot10^6$ K and $B_0=0.06$ T. The magnetic field value is based on the observations by \citet{ReinersBasri10} and the coronal temperature is a typically applied temperature in the literature. The mass flux of $\dot{M}=10^{10}\,\mathrm{kg}\,\mathrm{s}^{-1}$ is taken from the estimated maximum mass flux of Proxima Centauri \citep{WargelinDrake02}.

\begin{figure*}[t]
\plotone{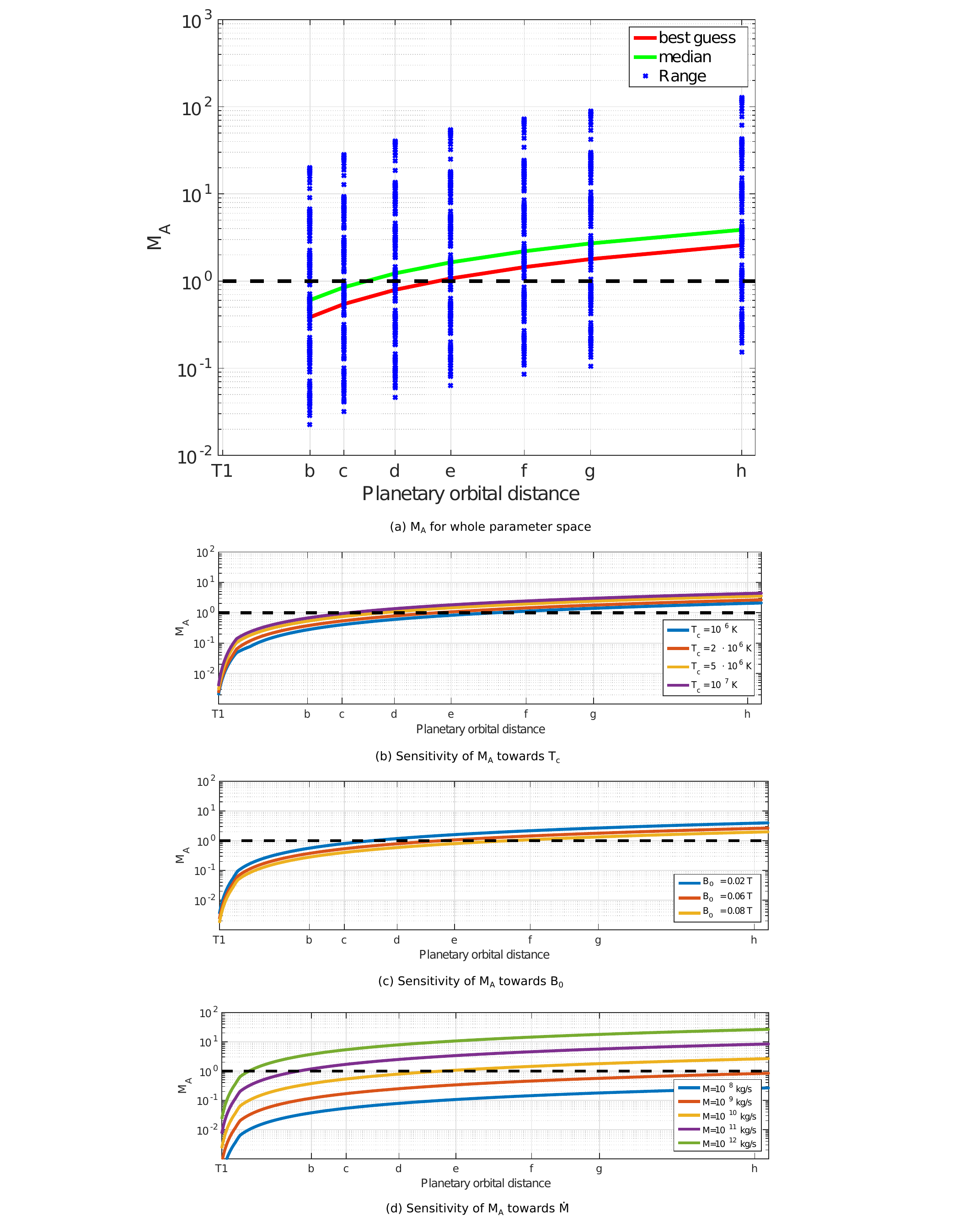}
\caption{Alfv\'{e}n Mach numbers $M_\mathrm{A}$ at each planet. Figure (a) shows a parameter study for $M_\mathrm{A}$ as a function of distance, indicated by the planetary orbital distance. Blue points indicate the local Alfv\'{e}n Mach number for every possible combination of the parameter space. The red curve shows the best guess result and the green curve represents the median of all points. Figures (b) to (d) present sensitivity studies for the coronal temperature $T_\mathrm{c}$, the equatorial stellar magnetic field strength $B_0$ and the mass flux $\dot{M}$. The respective other parameters are assigned to their best-guess values. \label{fig:Machnumbers}}
\end{figure*}

\subsection{Mach numbers and Alfv\'{e}n wings}\label{sec:Machnumberssection} 
The Alfv\'{e}n Mach number at each planet determines the planets ability to form an Alfv\'{e}n wing and connect to the star. Our chosen parameter space leads to a range of $M_\mathrm{A}$ at the planets as displayed in figure \ref{fig:Machnumbers} (a). The blue dots represent our estimated combinations of $\dot{M}$, $B_0$ and $T_c$, which we scanned logarithmically equally spaced within the limits given in section \ref{sec:parameterspace}. Our 'best-guess' is shown in red (see section \ref{sec:parameterspace}). In comparison the green line shows the median of all $M_\mathrm{A}$, which can be used as an indicator of the probability that a planet experiences $M_\mathrm{A}<1$ within our assumed parameter space. Table \ref{tab:SPIresults} shows the basic quantities of the TRAPPIST-1 system that are required for our analysis. The last two columns present our best-guess results of the Mach numbers $M_\mathrm{A}$ with minimum and maximum values in sub- and superscripts and the same for the Poynting flux.

It is visible that within our parameter space sub- and super-Alfv\'{e}nic conditions are possible for all planets. For T1b and c the best-guess indicates $M_\mathrm{A}$ of $~0.4$ and $~0.6$ (table \ref{tab:SPIresults} and figure \ref{fig:Machnumbers} (a)) while the maximum lies at $M_\mathrm{A}=20$ and $28$ and the minimum at $0.02$ and $0.03$. The median lies around 0.6 and 0.9 respectively. This implies that T1b and c are more likely to be exposed to sub-Alfv\'{e}nic flows than not and thus likely couple via Alfv\'{e}n wings to the star. T1d has $M_\mathrm{A}=0.8$ in the best-guess assumption (table \ref{tab:SPIresults}) and the median lies at $M_\mathrm{A}=1.1$, which makes it questionable if it constantly experiences sub-Alfv\'{e}nic conditions. However due to variations in the stellar wind environment it might occasionally be possible for T1d to form Alfv\'{e}n wings. The other planets are unlikely to exhibit SPI because best-guess and median lie above $M_\mathrm{A}=1$ (table \ref{tab:SPIresults} and figure \ref{fig:Machnumbers} (a)).

Figures \ref{fig:Machnumbers} (b) to (d) show the dependence of the Mach number on the three model parameters. We vary only one parameter at a time within our defined parameter space. Subfigure (b) shows the Mach number in dependence of the coronal temperature $T_\mathrm{c}$, similarly in subfigures (c) and (d) for $B_0$ and $\dot{M}$. For each sensitivity study the other parameters are kept constant to their respective best-guess values.

We see in figure \ref{fig:Machnumbers} (b) that the blue curve for $10^6$ K represents the minimum curve, while for temperatures of $10^7$ K (purple) the Mach numbers are larger due to the higher stellar wind velocities. The span between the resulting Mach numbers is about half an order of magnitude and thus smaller than the parameter spread of $T_\mathrm{c}$ itself. This variability is smaller than the variability of three orders of magnitude obtained in the Mach numbers in figure \ref{fig:Machnumbers} (a). We obtain similar results from the magnetic field study in figure \ref{fig:Machnumbers} (c). $B_0$ is known within a small range of uncertainty compared to the other two parameters. The blue curve represents a stellar magnetic field of $0.02$ T while the red and yellow curves represent $0.06$ and $0.08$ T. With less than half an order of magnitude the minimum-maximum spread of the obtained Mach numbers can not account for the spread of almost three orders of magnitude calculated with the whole parameter space. In the studies of $T_\mathrm{c}$ and $B_0$ the planets T1b and T1c always lie within the sub-Alfv\'{e}nic range whereas T1e is the outermost candidate for SPI.

The mass flux (figure \ref{fig:Machnumbers} (d)) is the least constrained quantity in our study and spans a range of four orders of magnitude between $10^8-10^{12}\,\mathrm{kg}\,\mathrm{s}^{-1}$. We see that smaller mass fluxes cause smaller Mach numbers than large mass fluxes, which is due to its dependence on the plasma particle density. Therefore low mass fluxes result in lower Mach numbers. The resulting Mach numbers show a spread of two orders of magnitude. Therefore the uncertainty of the mass flux is responsible for the obtained large spread in the Mach numbers. As one can see in the red, yellow and purple curves the difference of one order of magnitude in the mass flux compared to the best-guess (yellow) can lead to totally different results concerning SPI. For weaker mass fluxes all planets may cause SPI but with mass fluxes stronger than $10^{10}\,\mathrm{kg}\,\mathrm{s}^{-1}$ there might be no SPI at all.

\begin{figure*}[t]
\plotone{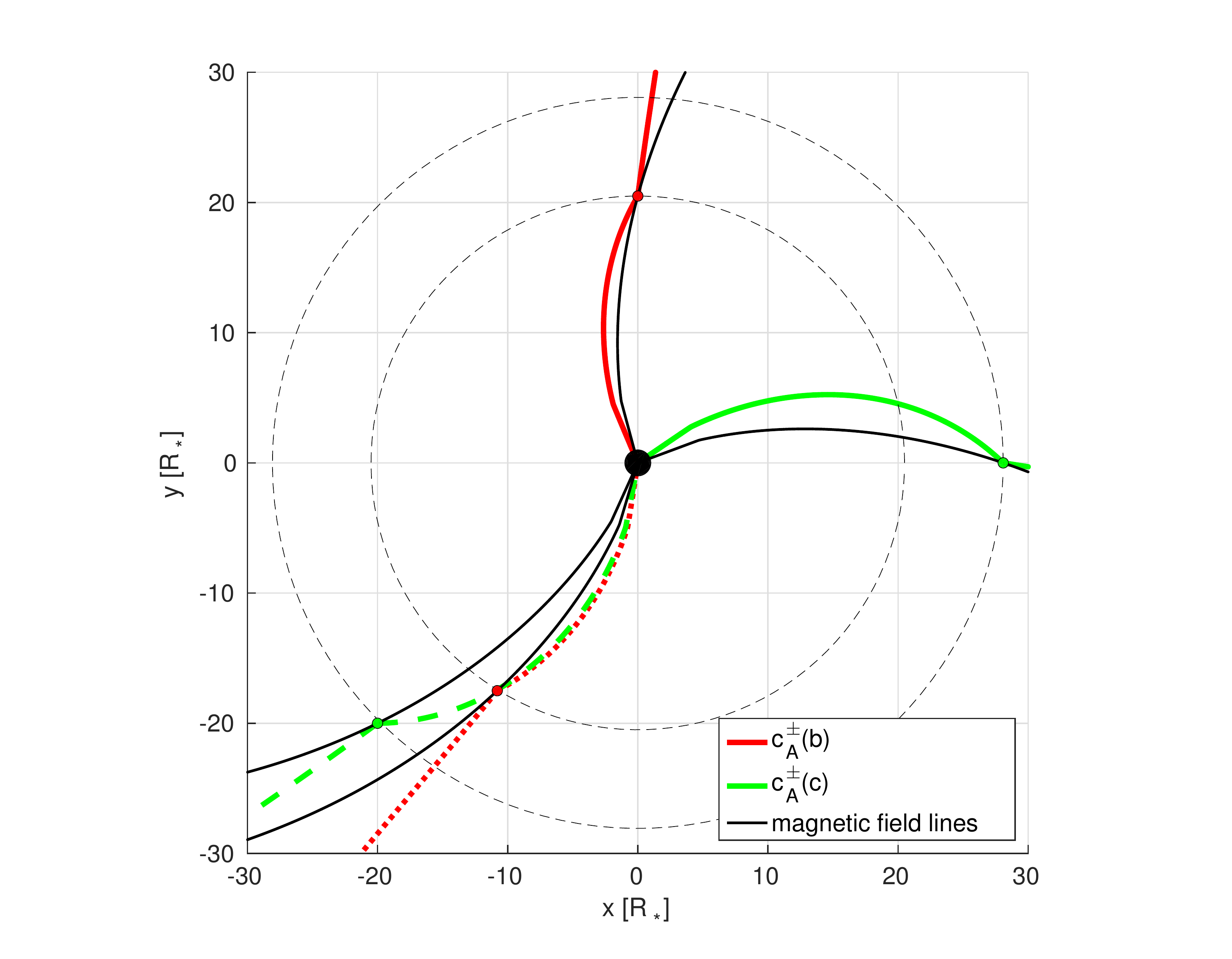}
\caption{Shown are the Alfv\'{e}n wings of T1b (red) and T1c (green). The plot shows two different situations: One with wings separated from each other (solid lines) and one with interacting wings (dashed lines). Additionally it shows in black the field lines that connect to the planets. \label{fig:Alfvenwings}}
\end{figure*}

The Alfv\'{e}n characteristics, i.e., the travel paths of the Alfv\'{e}n waves, are given by
\begin{equation}
   \textit{\textbf{c}}_\mathrm{A}^\pm=\textit{\textbf{v}}_\mathrm{rel}\pm\textit{\textbf{v}}_\mathrm{A}\label{eq:Alfvencharacteristic}
\end{equation}
according to \citet{Neubauer80} and \citet{Saurea13}. Figure \ref{fig:Alfvenwings} shows the suggested location of the associated Alfv\'{e}n wings. The red and green lines show $c_\mathrm{A}^\pm$ of T1b  and T1c respectively. The black lines represent field lines which connect the star to the planets. Here we also see that one of the respective Alfv\'{e}n wings is directed towards the star. Figure \ref{fig:Alfvenwings} presents two situations: The solid lines, where the Alfv\'{e}n wings separately interact with the star, and another time, represented by dashed lines, where the wings from the two planets merge into each other.

\citet{Garraffoea17} conclude from their simulations that all planets from T1b to T1g reside within the Alfv\'{e}n shell for large parts of their orbits. Mainly this is due to a smaller mass flux in their model. If we apply $\dot{M}=2\cdot10^9\,\mathrm{kg}\,\mathrm{s}^{-1}$ in our best guess, we gain a similar result as \citet{Garraffoea17}. In that case all planets except T1h are within the Alfv\'{e}n radius as well.
A good indicator for the quality of our model compared to the MHD simulations is the dynamic pressure of the solar wind at each planet. \citet{Garraffoea17} investigate the space weather of the planets to find out how it might affect their habitability. The dynamic pressure is an important parameter in these estimations because it indicates the influence of the stellar wind on a planets atmosphere. We estimated the dynamic pressure for a similar set of parameters as \citet{Garraffoea17}. The largest deviations appear at T1b with $40\%$ and the smallest at T1e with $5\%$. On average the relative deviation lies at $18\%$, which is small considering the uncertainties in the input parameters and the general differences in the applied models.

\begin{figure*}[t]
\plotone{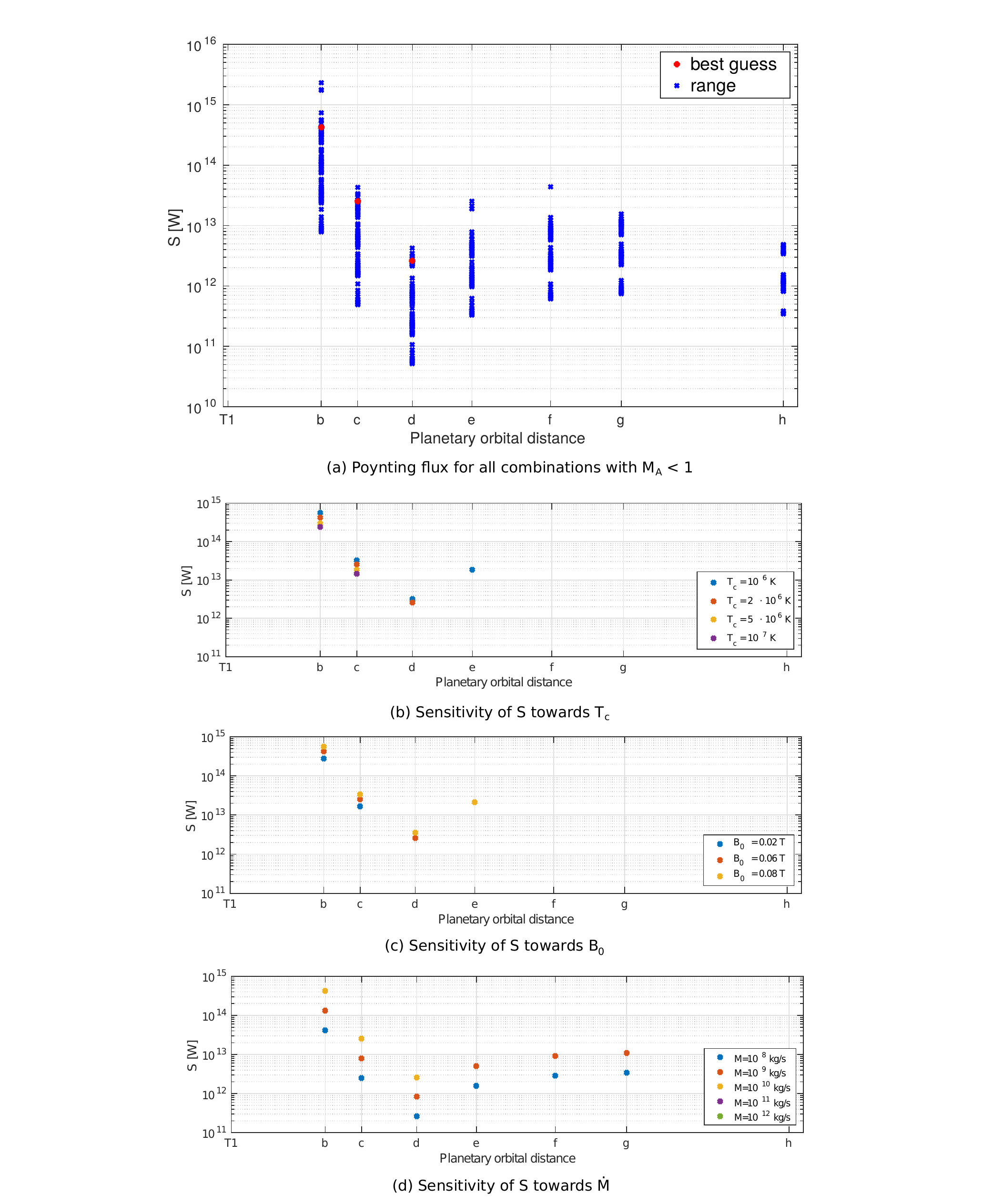}
\caption{Poynting fluxes generated by the planets. Figure (a) shows the Poynting flux as a function of distance for all parameter combinations that allow $M_\mathrm{A}<1$ (blue dots) with the best-guess in red again. Figures (b) to (d) present sensitivity studies for each of the three model parameters the coronal temperature $T_\mathrm{c}$, the equatorial magnetic field strength $B_0$ and the mass flux $\dot{M}$. The respective other parameters are assigned to their best-guess values. \label{fig:Poyntingfluxes}}
\end{figure*}

\subsection{Poynting fluxes}\label{sec:Poyntingfluxes}
The next step is to estimate the Poynting flux $S$ for all planets as shown in figure \ref{fig:Poyntingfluxes} (a). Table \ref{tab:SPIresults} summarises the Poynting fluxes calculated with the best-guess assumption. The superscripted numbers indicate the maximum Poynting flux that is possible with our parameter space. No lower values are given because our parameter space also allows non-existing Alfv\'{e}n wings for each planet. We first assume that the planets do not possess intrinsic magnetic fields and no expanded atmospheres. For T1b and T1c the Poynting flux then lies between $5\cdot10^{11}$ W and $10^{15}$ W. The best-guess indicates powers of $4.2\cdot10^{14}$ W for T1b and $2.5\cdot10^{13}$ W for T1c (table \ref{tab:SPIresults}). 
For some parameters the planets T1d to T1h experience $M_\mathrm{A}<1$ and generate Poynting fluxes between $5\cdot10^{10}$ and $3\cdot10^{14}$ W. Our best-guess gives $S=0$ W because for these parameters the planets experience $M_\mathrm{A}>1$. The Poynting flux generated by an exoplanet is strongly influenced by the angle $\Theta$ between the relative plasma velocity $\textbf{\textit{v}}_\mathrm{rel}$ and the normal of the stellar magnetic field $\textbf{\textit{B}}$ \citep{Saurea13}. If this angle is $90^\circ$ the magnetic field and the relative flow direction are parallel to each other. Therefore there is approximately no disturbance perpendicular to the magnetic field which causes an Alfv\'{e}n wave and accordingly there is no Poynting flux generated by the planet. TRAPPIST-1 rotates with a period of $3.3$ d so the point of field-parallel flow lies between T1c and T1d. Thus T1b generates the strongest Poynting flux and T1d generates the weakest Poynting fluxes because it lies very close to this point of field-parallel flow. However T1d is also the second smallest planet in the system, which also causes a weaker Poynting flux than its neighbours.

Figures \ref{fig:Poyntingfluxes} (b) to (d) show the Poynting flux in dependence of the three model parameters $T_\mathrm{c}$, $B_0$ and $\dot{M}$ for all cases where $M_\mathrm{A}<1$. This allows us to discuss the sensitivity of the Poynting flux on the different parameters similar to section \ref{sec:Machnumberssection}.

In subfigure \ref{fig:Poyntingfluxes} (b) the red dots indicate the best-guess with $T_\mathrm{c}=2\cdot10^6$ K, yellow and purple indicate higher temperatures and the blue dots represent $10^6$ K. The temperature affects the Poynting flux via the stellar wind velocity. According to equation \ref{eq:poyntingformula} one expects larger Poynting fluxes for higher coronal temperatures. However in our study lower temperatures lead to slightly larger Poynting fluxes. This is due to the additional effects of the geometry between stellar wind and stellar magnetic field, represented by the angle $\Theta$. TRAPPIST-1's planets are within a region close to the star where the flow is nearly aligned with the magnetic field lines. For smaller stellar wind velocities (and $T_\mathrm{c}$) the flow is less aligned with the magnetic field and causes stronger magnetic field perturbations resulting in stronger Poynting fluxes. At T1b one can see that the spread between minimum and maximum is about half an order of magnitude from $2\cdot10^{14}$ W to $6\cdot10^{14}$ W. The effect is similar at T1c with $1.5\cdot10^{13}$ W to $5\cdot10^{14}$ W. The spread therefore lies around half an order of magnitude and is similar to the spread of the Alfv\'{e}n Mach number.

Figure \ref{fig:Poyntingfluxes} (c) shows the dependence of the Poynting flux on the stellar magnetic field strength $B_0$. At T1b the Poynting flux spreads from $3\cdot10^{14}$ W to $6\cdot10^{14}$ W. At T1c the Poynting flux is lower and ranges from $1.8\cdot10^{13}$ W to $3.5\cdot10^{13}$ W. This small spread also is similar to the obtained Mach numbers and the narrow parameter range of $B_0$.

We have seen that the chosen parameter range of the mass flux has the strongest impact on the Mach numbers. However only the mass fluxes between $10^8-10^{10}\,\mathrm{kg}\,\mathrm{s}^{-1}$ allow the existence of SPI. The best-guess value of $10^{10}\,\mathrm{kg}\,\mathrm{s}^{-1}$ (yellow) allows three planets to generate SPI and the lower mass fluxes (red and blue) would enable all planets to generate SPI. Due to the lower associated plasma densities the Poynting flux depends proportionally on the mass flux. The minimum-maximum spread is about one order of magnitude at all three inner planets. This agrees well with the spread in the Mach numbers for the same parameter range which is also about one order of magnitude.

\begin{figure*}[t]
\plotone{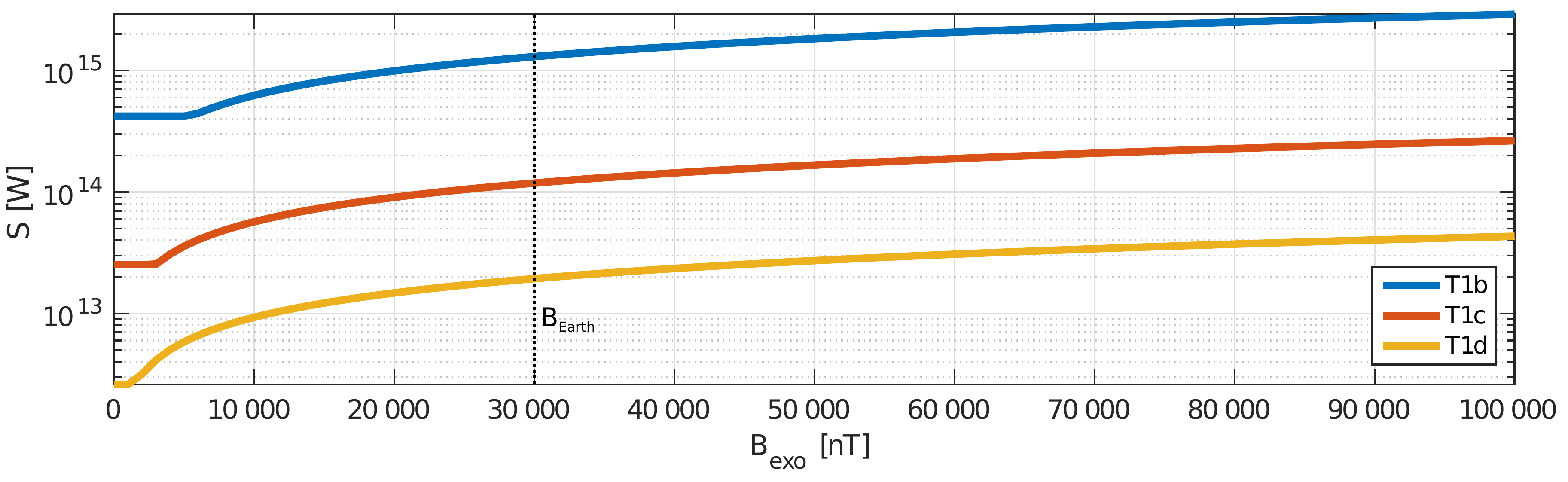}
\caption{Poynting fluxes generated by the three innermost planets assuming they have intrinsic magnetic fields. The Poynting flux $S$ is shown as a function of the intrinsic equatorial field strength $B_\mathrm{exo}$. The different line colors represent the planets: blue for T1b, red for T1c and yellow for T1d. The equatorial magnetic field strength of the Earth $B_\mathrm{Earth}$ is highlighted by the black dashed line. \label{fig:Bexo}}
\end{figure*}

In case the planets possess intrinsic magnetic fields the obstacle size responsible for SPI increases. To estimate this size we use the expression for the effective size of the magnetic obstacle 
\begin{equation}
R_\mathrm{eff}= k \, R_\mathrm{obst} = k \, R_\mathrm{p}\,\left(\frac{B_\mathrm{exo}}{B_\mathrm{loc}}\right)^\frac{1}{3}
\end{equation}
as given by equation (57) in \citet{Saurea13}, where $R_\mathrm{obst}$ characterizes the size of the closed magnetic field line region of the planet, $R_\mathrm{p}$ the planetary radius and $k$ is a factor that determines the effective size based on the geometry of the planetary magnetic field (see expression in Figure 7 in \citet{Saurea13} or discussion in \citet{Strugarek16}). Here we assume that the intrinsic dipole moment and the stellar magnetic field are parallel, which leads to the maximum effective size of the Alfv\'{e}n wing of $R_\mathrm{eff}=\sqrt{3}\,R_\mathrm{obst}$ with $k=\sqrt{3}$.

Figure \ref{fig:Bexo} shows how the Poynting flux depends on the intrinsic planetary magnetic field strength. We apply the best-guess to characterise the stellar wind and calculate the Poynting fluxes generated by the three innermost planets with their respective assumed magnetospheres. We chose equatorial magnetic field strengths between $0$ nT, i.e., no magnetosphere at all, up to $100\,000$ nT. This range includes the strength of the terrestrial magnetic field of about $B_\mathrm{Earth}=30\,000$ nT (black dashed line in figure \ref{fig:Bexo}). Due to the strong stellar magnetic field and the resultant large stellar wind magnetic field near the orbital distances of the planets, the effective sizes may still not be larger than the planet itself. We took this into account and see that T1b starts to develop closed magnetospheric field lines with a surface field strength of $~5000$ nT, visible in the curve when the Poynting flux starts to deviate from the value of $4.2\cdot10^{14}$ W. For an earth-like magnetic field T1b produces $1.5\cdot10^{15}$ W. For stronger fields of up to $100\,000$ nT the expected Poynting fluxes increase by a factor of $~2$ towards $~3\cdot10^{15}$ W. The same pattern applies for the other two planets, both show an increase of one order of magnitude for intrinsic magnetic fields up to $100\,000$ nT.

\citet{Turnpenneyea18} carried out a study on the radio luminosity of SPI in several systems, including TRAPPIST-1. They included a parameter space of three hypothetical intrinsic magnetic field strengths based on the earth's magnetic field and found that the upper limits of the estimated radio luminosity may be observable for future radio telescopes. The authors apply stellar wind parameters with a solar mass flux of $2\cdot10^9\,\mathrm{kg}\,\mathrm{s}^{-1}$ and therefore all planets generate SPI. For the terrestrial field strength their model generates a magnetosphere with closed field lines only for the outer planets starting at T1d. Our model includes a range up to the distance $r_2=5\,R_*$ where the magnetic field behaves dipolar whereas \citet{Turnpenneyea18} applied the Parker field configuration right from the star. Therefore our radial magnetic field strength is weaker and allows larger planetary magnetospheres. This shows that the stellar magnetic field geometry plays an important role in these types of studies.
\citet{Garraffoea17} also investigated the case of a $50\,000$ nT field strength for all planets and estimated that T1b would have a magnetopause distance of approximately $1$ to $1.5\,R_\mathrm{p}$ and T1c of approximately $1.2$ to $1.5\,R_\mathrm{p}$. Our calculated obstacle sizes are similar with values of $R_\mathrm{obst}=1.2\,R_\mathrm{p}$ for T1b and $1.5\,R_\mathrm{p}$ for T1c. This shows that our model results are similar to what can be estimated by numerical simulations but our study allows the scan of a larger parameter space, which is not possible with computationally expensive MHD simulations.

\begin{figure*}[t]
\plotone{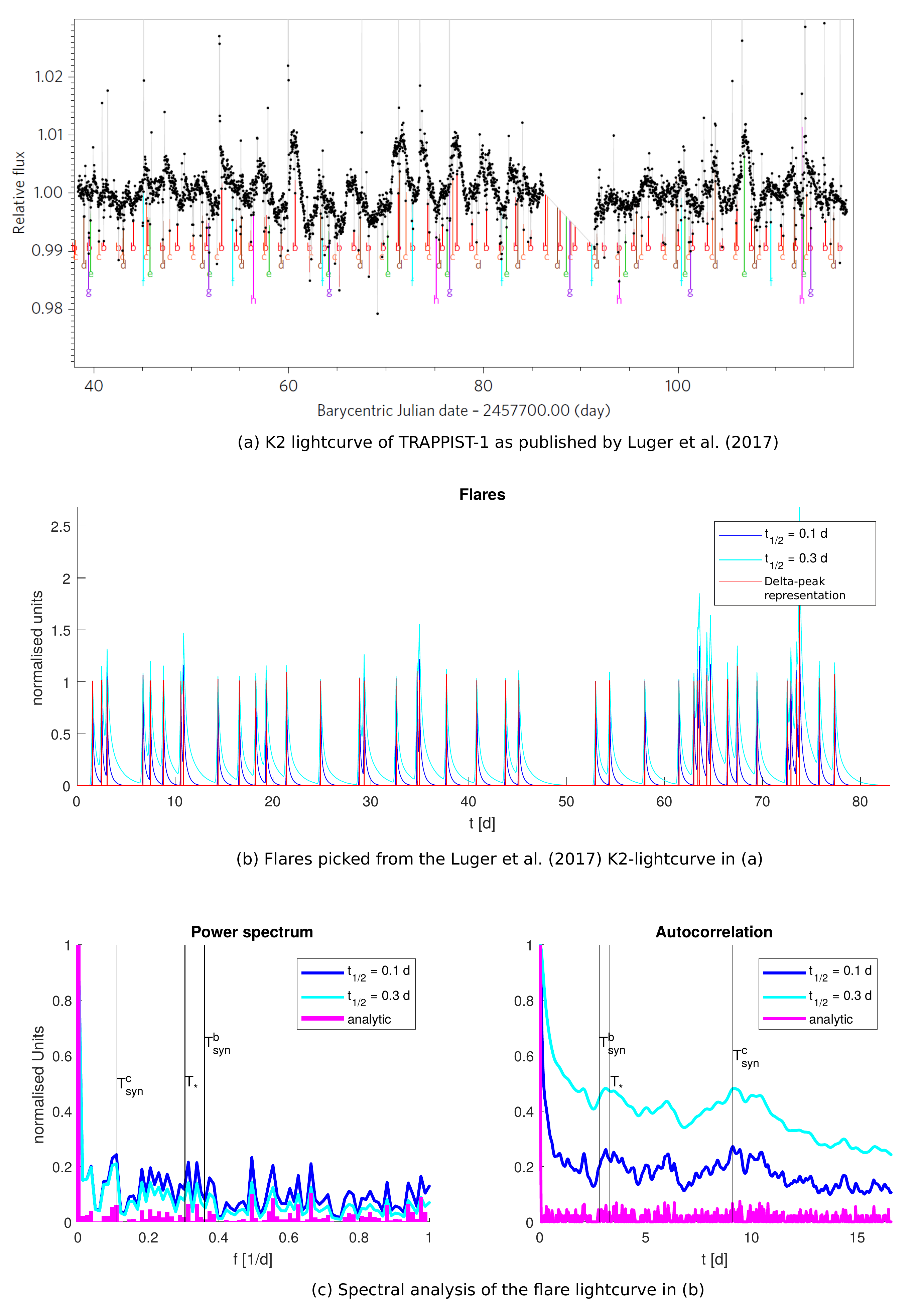}
\caption{Analysis of TRAPPIST-1's flare time series. Panel (a) presents the reprinted figure 2 (b) from \citet{Lugerea17} to illustrate the shape of the lightcurve and the appearance of flares. Panel (b) shows the flares that we picked by eye from the lightcurve in (a) in red. We represented the flares with delta-peaks and additionally with the empircal flare template by \citet{Davenportea14} for two different widths. Panel (c) shows an analysis of the flare lightcurve. The left plot presents the analytical power spectrum of the delta peak representation (in magnenta) and the numrical power spectrum of the empirical flare representation after \citet{Davenportea14} (blue and cyan). The right plot shows the autocorrelation of the time series.\label{fig:FlarePeriodicity}}
\end{figure*}

\subsection{Temporal variability in the system} 
In this section we investigate if SPI might be visible in the lightcurves of TRAPPIST-1. In the first part we study if SPI generates detectable signals. In the second part we will use the 78 day long K2-lightcurve of TRAPPIST-1 and try to identify quasi-periodic signals possibly associated with T1b and T1c that we predicted with our analytic considerations of temporal variability. In the third part we discuss briefly how the height of a flare affects its visibility and therefore a periodicity of the flare lightcurve.

\subsubsection{Is the signal detectable?}
TRAPPIST-1 has a bolometric luminosity of approximately $2\cdot10^{25}\ \mathrm{W}$ \citep{Grootelea18}, which corresponds to an energy flux density of $2.4\cdot10^{8}\ \mathrm{W}\,\mathrm{m}^{-2}$ on its surface. Assuming that T1b possesses an intrinsic magnetic field with $30\,000$ nT its Alfv\'{e}n wing carries an energy flux of $1.3\cdot10^{15}\ \mathrm{W}$ in the best-guess assumption. Using $\nabla \cdot \textbf{\textit{B}} = 0$ we can estimate the approximate size of the Alfv\'{e}n wing's footprint and find an energy flux density of $10^{4}\ \mathrm{W}\,\mathrm{m}^{-2}$ at the surface of the star. This flux density is about $10^{-4}$ of the stellar background. Temporal variability due to the plain visibility of the Alfv\'{e}n wing's footpoint or through a tilted stellar dipole field is thus expected to be hardly observable in broadband photometric lightcurves.

In this study we take the Poynting flux as the upper limit for the incoming power at the star and also the expected emissions. Alfv\'{e}n waves in sufficiently smooth magnetic fields are even in the non-linear case neither dispersive nor dissipative \citep{Neubauer80}. Details of the propagation properties however depend on the mostly unknown properties of the stellar winds.

Observations at X-ray wavelengths conducted by \citet{Wheatleyea17} revealed an X-ray luminosity of $L_\mathrm{X}/L_\mathrm{bol}=2-4\cdot10^{-4}$. Observations with the Hubble Space Telescope (HST) by \citet{Bourrierea17} showed that TRAPPIST-1's Lyman-$\alpha$ emissions are about three times weaker than its X-ray luminosity. According to \citet{Bourrierea17} TRAPPIST-1's chromosphere is only moderately active compared to its corona and transition region. The corresponding flux density in X-ray and Ly$\alpha$ is therefore of the same order of magnitude as our estimate for T1b's energy flux density. This serves as an example that the detectability of SPI is strongly enhanced if the emission associated to it is concentrated in certain wavelength bands only. However both of these observations are, at $7.8$ hours in the X-ray \citep{Wheatleyea17} and four HST orbits \citep{Bourrierea17}, too short to identify temporally variable planetary signals.

\subsubsection{Search for quasi-periodicity in TRAPPIST-1's flares}\label{sec:SynodicsignalsT1}
Another promising approach to search for SPI via photometry is to look for quasi-periodic signals in the flares of TRAPPIST-1. This assumes that Alfv\'{e}n wings trigger flares on the star as discussed in section \ref{sec:TriggerAnomaly}. 

\citet{Lugerea17} published TRAPPIST-1's K2-lightcurve, that is corrected for systematic errors and has low-frequency trends removed. Figure \ref{fig:FlarePeriodicity} (a) shows a replot of this lightcurve from figure 2 in \citet{Lugerea17}. The stellar variability is well visible and flares appear as sudden increases in luminosity and sometimes as single spikes \citep{Lugerea17}. For our analysis we read out the times and relative fluxes of flare events by eye (shown as red spikes in figure \ref{fig:FlarePeriodicity} (b) upper panel). Our sample of 41 flares coincides with the sample from \citet{Vidaea17}, who analysed 42 flares by eye as well. \citet{Vidaea17} estimated typical integrated flare energies of $10^{23}\,-\,10^{26}$ J whereas only one flare reached a total energy of $10^{26}$ J in this lightcurve. The bulk of the flares has much lower energies of $10^{23}$ to $10^{24}$ J. Thus the observed flares fit well into the theoretical flare energy expectations from \citet{Lanza18}. According to this theory we calculate a range of maximum flare energies of $~3\cdot10^{24}$ J to $~2\cdot10^{27}$ J for TRAPPIST-1.

To find possibly existing periods among the flares we perform a spectral analysis of the flare lightcurve. We approximate the flares as Delta-peaks via
\begin{equation}
 f(t) = \sum_{\mathrm{n}=1}^N A_\mathrm{n}\,\delta(t-t_\mathrm{n})\label{eq:FlareDeltaApprox}
\end{equation}
with $N$ the total number of flares, $t_\mathrm{n}$ the time when each flare occurs and $A_\mathrm{n}$ its amplitude. We alternatively represent the flares with a finite duration through the empirical flare template by \citet{Davenportea14} given by their equations (1) and (4). \citet{Vidaea17} and \citet{Vidaea18} for example apply this shape for their analysis of TRAPPIST-1's flares. This flare template describes the shape of a flare with a sharp rise phase and a decay phase after the main event in terms of the full time width at half the maximum $t_{1/2}$. For our analysis of the flares we apply two different $t_{1/2}$ (figure \ref{fig:FlarePeriodicity} upper plot), one narrow representation with $t_{1/2}=0.1$ d (blue) and one broad representation with $t_{1/2}=0.3$ d (cyan).

In our analysis we compute the power spectrum and the autocorrelation of TRAPPIST-1's flares. For the delta-peak representation from equation \ref{eq:FlareDeltaApprox} we calculate the power spectrum from the Fourier series (see appendix \ref{sec:AppFourierseries}) and for the template-fitted flares we compute the power spectrum from numerical FFT. Additionally we calculated the autocorrelation function (see appendix \ref{sec:AppFourierseries}) 
\begin{equation}
R(\tau) = \sum_{\mathrm{i}=1}^N \sum_{\mathrm{j}=1}^N A_\mathrm{i}\,A_\mathrm{j}\, \delta(\tau-t_\mathrm{i}+t_\mathrm{j})
\end{equation}
from the delta-peaks analytically, with the time lag $\tau$. We also numerically calculate the autocorrelation from the continuous template-fitted flares.

The bottom left plot of figure \ref{fig:FlarePeriodicity} shows the respective power spectra of the delta-peak time series in magenta and of the empirical representation of the flares in blue and cyan. All power spectra are arbitrarily normalised because we are only interested in identifying certain frequencies. We show the spectral area of interest between $f=0$ and $1\ \mathrm{d}^{-1}$ and highlight the rotation period of TRAPPIST-1 $T_*$ and the expected synodic periods of T1b and T1c. We see a local maximum at the synodic period of T1c. However this maximum is not significantly larger than other local maxima in this range. There is also a slightly offset maximum at the stellar rotation period but no indication for a signal that is connected to T1b. 

In the bottom right plot we see the autocorrelation of the template-fitted flares in blue and cyan together with the analytic autocorrelation of the delta-peak flare light curve (magenta). To resolve the synodic period of T1c, we computed the autocorrelation for lags of up to $0.2\,t_\mathrm{max}$. The analytic autocorrelation shows a clustering of correlation peaks around the synodic period of T1c and the stellar rotation period $T_*$. For the blue template-fitted flares the correlation is generally about 0.1 with local maxima of above 0.2 at the highlighted periods. A broadening of the flares makes the autocorrelation more tolerable to small variations in a quasi-periodic occurrence of flares and shows a stronger correlation among the flares. Therefore the autocorrelation of the broad fitting curves (cyan) shows hints for a periodicity of around 9 d, which would correspond to a signal that appears with T1c's synodic period. The corresponding correlation lies around 0.5. The interval around the stellar rotation period of 3.3 d and T1b's synodic period of 2.78 d also shows a local maximum of about 0.5 in the cyan curve. However it is uncertain if this peak indicates a signal related to T1b or to the stellar rotation, although similar to the power spectrum it shows a tendency towards $T_*$. All in all, despite the hints, it remains uncertain if the flare lightcurve includes signals from SPI. 

\subsubsection{Visibility of triggered flares at TRAPPIST-1}\label{sec:VisibilityFlareT1}
Observations of flares are also subject to the visibility concern introduced in section \ref{sec:TriggerAnomaly}. If flares have negligible height, they are only visible 50\% of the time.
Reliable flare height estimations are only available for solar flares. Early studies from \citet{Warwick55} and \citet{WarwickWood59} about flare height distributions showed that most solar flares have low altitudes of a few thousand kilometres. A few very large flares appeared to have heights of $20\,000$ to $50\,000$ km \citep{WarwickWood59}. Observations of a solar limb flare by \citet{BattagliaKontar11} showed that white light originated from heights of $1500$ to $3000$ km, hard X-rays from $800$ to $1700$ km and EUV emissions from around $3000$ km. \citet{Pillitteriea10} observed a flare on the K-type star HD 189733 and estimated the furthest extent of the flaring loop to be on the order of the stellar radius (0.76 solar radii). These studies show that flares appear with a large variety of sizes but the heights of flares on M-dwarfs like TRAPPIST-1 are currently unknown. In our study we assume typical sun-like flare heights in a range between $1000$ km and $10\,000$ km. Those heights are non-negligible compared to the small radius of TRAPPIST-1 and can considerably extend the visible area.

\begin{figure*}
\plotone{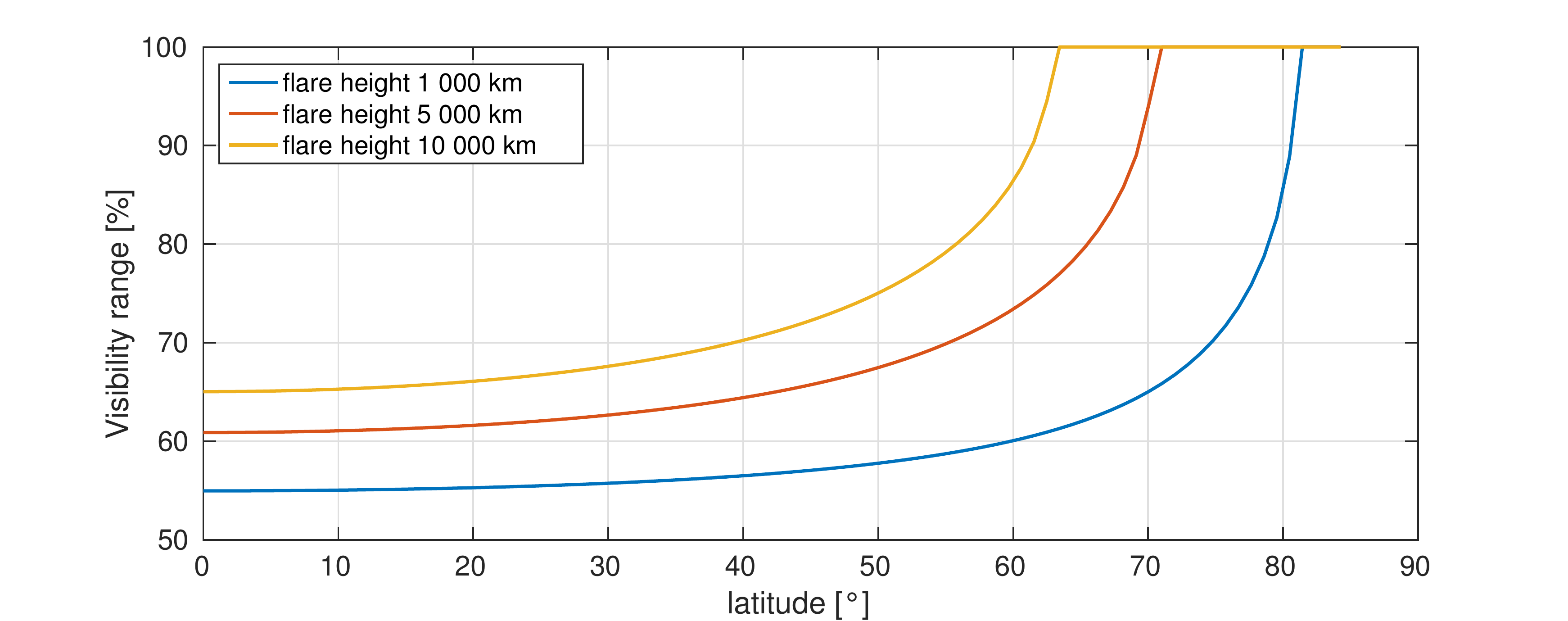}
\caption{This figure shows how the visibility of flares, in \% of the stellar surface, depends on flare height and latitude. The colours depict different flare heights.\label{fig:ExtendedVisibilityResults}}
\end{figure*}

Figure \ref{fig:ExtendedVisibility} and equation \ref{eq:VisExtensionFormula} show that the visibilty increases for various flare heights depending on their latitudinal location. In figure \ref{fig:ExtendedVisibilityResults} (b) we show how much the flare visibility is extended at TRAPPIST-1. The visibility range is presented as a function of latitude and in different colours for different flare heights $\Delta r$. TRAPPIST-1 has a radius of $R_*=81342$ km \citep{Gillonea17}. The blue curve shows $\Delta r=1\,000$ km, the orange one $\Delta r=5\,000$ km and the yellow curve $\Delta r=10\,000$ km. The sizes are chosen in accordance with the discussion in section \ref{sec:TriggerAnomaly}. They represent typical sizes of solar flare white light emissions. In this study we assume these flare heights for TRAPPIST-1. We see that the visibility of a flare at the equator is extended to 55\% for $\Delta r=1\,000$ km and to 65\% for $10\,000$ km. All three curves increase towards a 100\% visibility with increasing latitude. Large flares with a height of $10\,000$ km reach a 100\% visibility at a latitude of $~65^\circ$. The investigated smaller flares reach this threshold at $~70^\circ$ and $~80^\circ$.

For the analysis of flare observations we have to know where the footpoint of the planetary Alfv\'{e}n wings are located. As a simple estimate we apply a magnetic dipole model and use the field line equation to map the field lines connecting the planets with the star given by $\cos^2(\lambda)=R_*/a$ with the latitude $\lambda$. We receive footpoint latitudes of $\lambda_\mathrm{b}=77^\circ$ and $\lambda_\mathrm{c}=79^\circ$ for T1b and T1c respectively. Therefore the visibility is likely significantly enhanced for a planetary triggered flare at high or mid-latitudes and lies around $70\%$ to $100\%$ for the assumed flare heights. The planets likely reside on quasi-open field lines and thus the Alfv\'{e}n wings' footpoints probably lie even closer to the poles.

As shown in section \ref{sec:SynodicsignalsT1} we only see hints for SPI generated by T1c but not by the other planets. Generally we would expect a clearer signal from T1b because its Poynting flux is one order of magnitude larger than the one from T1c and additionally its synodic period of $2.78$ d is much shorter compared to the period of $9.1$ d of T1c. However, in case that the Alfv\'{e}n wing acts as a trigger of a flare it requires that a magnetic anomaly lies on its path across the star. Therefore an anomaly on the same latitude as the footpoint of the Alfv\'{e}n wing from T1c and no anomaly near T1b's footpoint could explain our findings. Star spots could be such magnetic anomalies that erupt flares. \citet{Rackhamea17} analysed the spot and faculae covering fractions of TRAPPIST-1 based on its rotational variability. The group estimated a spot covering fraction of 8\% with small spots and a faculae covering fraction of 54\%. On the sun the sunspots only occur at lower latitudes around the equator, which would probably not allow the Alfv\'{e}n wings to interact with these magnetic anomalies. However there is evidence that starspots and active regions on M-dwarfs occur on all latitudes \citep{BarnesandCollierCameron01}. This makes M-dwarfs like TRAPPIST-1 interesting targets to further look for this trigger mechanism because star spots might occur at higher latitudes as well, where SPI of planets more than several stellar radii away will couple to the star and can trigger flares that should be visible to an observer on Earth.

\section{Conclusions}
In the previous sections we discussed four mechanisms that lead to qualitatively different types of temporal variability of SPI in stellar lightcurves. Two of them are caused by orbital positions of the planets, namely the visibility and the wing-wing interaction. The other two mechanisms are due to the stellar magnetic field, either a tilted dipole field or coronal magnetic loops in active regions. The well known Jupiter-Io interaction appears with half the synodic rotation period of Jupiter as seen from Io due to Jupiter's tilted dipole magnetic field. Our study shows that temporal variability in SPI can be much more diverse than only the planetary orbital period. Instead we saw that the synodic and half the synodic period can play a major role in the time-variability of SPI. We chose the example of TRAPPIST-1 to look for temporal variability associated with SPI. We found from our semi-analytic parameter study that the two innermost planets probably generate SPI. Although the planets are expected to generate large Poynting fluxes, the generated signal is likely not visible within the broadband stellar background luminosity. Therefore we concentrate on a trigger mechanism where the planetary Alfv\'{e}n wing triggers the release of magnetic tension \citep{Lanza09,Lanza12,Lanza18}. We performed an analysis of TRAPPIST-1's flare time-series as observed by the K2-mission \citep{Lugerea17}. Our results hint at a quasi-periodic occurrence of flares with T1c's synodic period of 9.1 d and the stellar rotation period of 3.3 d but the results are inconclusive.

Our study applies a large parameter space for the stellar wind environment of TRAPPIST-1 to see whether its properties can cause SPI. In the corresponding sensitivity study in section \ref{sec:Poyntingfluxes} we find that the ability of planets to generate SPI is strongly affected by the uncertainty of the mass flux $\dot{M}$. The chosen parameter range of $\dot{M}=10^8-10^{12}\,\mathrm{kg}\,\mathrm{s}^{-1}$ caused a spread of the Alfv\'{e}n Mach number over two decades (fig. \ref{fig:Poyntingfluxes} (d)). Mass fluxes inferred from obserbations could allow a more precise determination of SPI properties. Even an upper limit of $\dot{M}$ will increase our understanding of the SPI in the system because it allows to put constraints on the minimum number of SPI generating planets.

In case SPI triggers flares on TRAPPIST-1, there are several reasons that prevent a clear identification in the flare lightcurves. The absence of a magnetic anomaly at certain stellar latitudes would cause flare triggering only for certain planets, e.g. possibly for T1c and not for T1b. If there is an anomaly at the right place there are still several physical reasons for irregularities in SPI-triggered flares that could affect a spectral analysis. The visibility of triggered flares causes a break in their periodicity as we have seen in section \ref{sec:VisibilityFlareT1}. It may also happen that not every interaction between an Alfv\'{e}n wing and the anomaly results in a flare. So the expected planetary signal might not occur every time. Furthermore magnetic anomalies may not be stationary, they could disappear just like sunspots while a new one forms somewhere else. This change results in a phaseshift of the flare time series. \citet{Lanza18} shows that for two of the presented mechanisms SPI can trigger flares, but flares may also erupt without the influence of a planet just as it happens on the sun. Finally flares and coronal mass ejections affect the stellar wind. Therefore the Alfv\'{e}n wings could experience a phase shift or could even be interrupted for a certain time until the stellar wind reaches its equilibrium state again. More confidence about a connection between flares and SPI can be established by long-time observations of flares, together with a mapping of magnetic anomalies on the corresponding stars. These cooperative observations would allow detailed analyses and might lead to a full confirmation of flare triggering SPI.

Flare occurence and thus detection is to some extend of stochastical nature, probably often influenced by the stellar rotation period, i.e. when an active region rotates into the visible hemisphere. We investigated the possibility of an external process that triggers flares periodically. For many stellar systems sufficiently long time-series might not be available to perform a robust spectral analysis to search for SPI triggered flare signals. A simpler indicator for SPI could be that stars with close-in SPI generating planets might on average have a higher flare occurence rate than stars without close-in planets.

We briefly discussed if SPI could be detectable in UV and X-ray and found that the observed energy flux densities are of a similar order of magnitude as planetary Poynting fluxes. \citet{Turnpenneyea18} modelled the expected radio luminosities of SPI at TRAPPIST-1 and concluded that future radio telescopes could be able to detect SPI-related radio emissions. A deep search for radio emission from the TRAPPIST-1 system by \citet{PinedaHallinan18} however could only derive upper limits. \citet{Franceea18} carried out a UV survey of 104 stars, with 71 planet-hosting stars, to characterise how the existence of planets affects the UV activity level of a star. The authors found that SPI provides no statistically significant contribution to the UV activity of their sample stars. Future modelling work could investigate which parts of the electromagnetic spectrum are excited by Alfv\'{e}n wings in stellar atmospheres. This would provide another possibility to help identify SPI in addition to temporal variability. 

Only out of curiosity, we like to mention that with the formerly considered rotation period of 1.4 days \citep{Gillonea16}, the synodic rotation period of T1c is 3.32 d and of T1b it is 19 d. T1c's synodic period is therefore nearly identical to the rotation period of 3.3 d obtained from the K2-lightcurve by \citet{Lugerea17} and \citet{Vidaea17}.

Electromagnetic star-planet interaction is a young area of research with only few and sometimes indirect observational evidences \citep{Shkolnikea03,Shkolnikea05,Scharf10,PoppenhaegerWolk14,Pillitteriea15}. Our study about temporal variability is motivated by the difficulty to observe SPI. Signals that show the aforementioned periods are therefore indicators for SPI. Stellar flares are clearly observable features and therefore the mechanisms presented in this paper could help to provide further evidence for SPI. In the K2-lightcurve we see hints for periodicities connected to electromagnetic SPI but the evidence is inconclusive. Recordings of longer lightcurves will provide an important basis to possibly identify SPI in star-planet systems and to understand the physical conditions in these systems.

\appendix

\section{Semi-analytic stellar wind model}\label{sec:AppStellarWindModel}
In this section we show the semi-analytical model, that we apply to describe TRAPPIST-1's stellar wind and magnetic field to calculate the properties of the expected SPI. We assume the stellar wind to be thermally driven and apply the model of \citet{Parker58} to calculate the stellar wind velocity $v_\mathrm{sw}$ based on the coronal temperature $T_\mathrm{c}$. This wind carries a total mass flux $\dot{M}$ in radial direction, given by

\begin{equation}
   \dot{M}=4\pi\,R_\mathrm{*}^2\,v_0\,\rho_0 \label{eq:massflux}
\end{equation}
with $v_0$ and $\rho_0$ the velocity and mass density in the corona. Thus, based on mass conservation, we can derive the radial density profile
\begin{equation}
    \rho(r)=\rho_0\,\left(\frac{R_\mathrm{*}}{r}\right)^2\,\frac{v_0}{v_\mathrm{sw}(r)}.\label{eq:densityprofile}
\end{equation}

We assume for simplicity that the stellar magnetic field can be described to a distance $r_2$ as a dipole field. For $r>r_2$ the field lines follow a Parker spiral \citep{Parker58} with
\begin{align}
   B_r&=B_0\left(\frac{R_\mathrm{*}}{r}\right)^2 \label{eq:ParkerMagneticField}\\
   B_\varphi&=-B_0\frac{\Omega_\mathrm{*}\,R_\mathrm{*}^2}{v_\mathrm{sw}\,r}.
\end{align}

\section{Fourier series of the K2 flare lightcurve}\label{sec:AppFourierseries}
In this section we present the Fourier series representation of a delta-flare lightcurve. We assume the $N$ flares to be delta-peaks with amplitude $A_\mathrm{n}$ represented by the function
\begin{equation}
f(t) = \sum_{\mathrm{n}=1}^N A_\mathrm{n}\, \delta(t-t_\mathrm{n})\label{eq:flarefunction}
\end{equation}
The function $f(t)$ can be represented by a Fourier series $f(t) = \sum_{\mathrm{k}=-\infty}^\infty \alpha_\mathrm{k} \, \exp \left( i \omega_\mathrm{k} t \right)$ with the Fourier coefficients $\alpha_\mathrm{k}$ and $\omega_\mathrm{k}=2 \pi k/T$ with the period $T$, which is assumed to be the length of the time-series $t_\mathrm{max}$ here. Those coefficients are given by
\begin{equation}
\alpha_\mathrm{k} = \frac{1}{t_\mathrm{max}} \sum_{\mathrm{n}=1}^N A_\mathrm{n} \exp \left( - i \omega_\mathrm{k} t_\mathrm{n} \right)
\end{equation}
Therefore the lightcurve can be represented by 
\begin{equation}
f(t) = \frac{1}{t_\mathrm{max}} \sum_{\mathrm{k}=-\infty}^\infty \sum_{\mathrm{n}=1}^N A_\mathrm{n} \exp \left( - i \omega_\mathrm{k} (t-t_\mathrm{n}) \right)
\end{equation}

To complete our analytic analysis of the delta-flare lightcurve we also calculate the autocorrelation of an arbitrary flare lightcurve.
We again apply equation \ref{eq:flarefunction} to represent our flare lightcurve via delta-peaks. The autocorrelation is defined by 
\begin{equation}
R(\tau) = \int_{-\infty}^{\infty} f(u+\tau)\, f(u) \, du
\end{equation}
and in our case it is given by
\begin{equation}
R(\tau) = \int_{-\infty}^{\infty} \left[ \sum_{\mathrm{i}=1}^N A_\mathrm{i}\, \delta(u+\tau-t_\mathrm{i}) \right] \left[ \sum_{\mathrm{j}=1}^N A_\mathrm{j}\,\delta(u-t_\mathrm{j})\right]du.
\end{equation}
We receive
\begin{equation}
R(\tau)  = \sum_{\mathrm{i}=1}^N \sum_{\mathrm{j}=1}^N A_\mathrm{i}\,A_\mathrm{j}\, \delta(\tau-t_\mathrm{i}+t_\mathrm{j}).
\end{equation}

\end{document}